\documentclass[12pt,journal,draftcls,a4paper,onecolumn]{IEEEtran} 
\usepackage{dsfont}

\usepackage{subeqnarray}
\usepackage{cleveref}
\usepackage{amssymb}
\usepackage{amsfonts}
\usepackage{amsmath}
\usepackage[dvipdfm]{graphicx}
\usepackage{bmpsize}
\usepackage{amsmath}
\usepackage[all,poly]{xy}
\usepackage{multirow}
\usepackage{algorithm}
\usepackage{algorithmic}
\usepackage{color}
\usepackage[noadjust]{cite}

\title{Spectral unmixing of Multispectral Lidar signals}
\author{Yoann Altmann, Andrew Wallace and Steve
McLaughlin\thanks{This study was supported by the Direction G\'en\'erale de l'armement, French Ministry of Defence and by EPSRC via grant EP/J015180/1.
}
\thanks{Yoann Altmann, Andrew Wallace and Steve
McLaughlin are with School of Engineering and Physical Sciences, Heriot-Watt University,
U.K. (email: \{y.altmann,a.wallace,s.mclaughlin\}@hw.ac.uk).}}
\newcommand{\bphi}{{\boldsymbol \phi}}

\def\bfb{{\mathbf{b}}}

\def\bfw{{\mathbf{w}}}

\def\bfJ{{\mathbf{J}}}

\def\bbR{{\mathbb{R}}}

\newcommand{\Vpix}[1]{\mathbf{y}_{#1}}

\newcommand{\MATpix}{\mathbf{Y}}
\newcommand{\pix}[2]{y_{#1,#2}}

\newcommand{\nbband}{L}

\newcommand{\nbbin}{T}
\newcommand{\nobin}{t}

\newcommand{\MATmat}{{\mathbf M}}
\newcommand{\Vmat}[1]{{\mathbf m}_{#1}}
\newcommand{\mat}[2]{m_{#1,#2}}

\newcommand{\paramvect}{\boldsymbol{\theta}}

\newcommand{\transp}{^T}

\newcommand{\Indicfun}[2]{\textbf{1}_{#1}\left(#2\right)}

\newenvironment{algogo}[1]{
\smallskip
\noindent \hrule\vspace{0.2\baselineskip} \hrule
\begin{small}
\refstepcounter{algo} \center{\bf \textsc{Algorithm \thealgo}}
\\{\center{\bf #1}}
\smallskip
\flushleft
 } {
\end{small}
\smallskip
\hrule\vspace{0.2\baselineskip} \hrule
\smallskip
}

\newcounter{algo}
\renewcommand{\thealgo}{\arabic{algo}}

\begin{document}
\maketitle

\begin{abstract}
In this paper, we present a Bayesian approach for spectral unmixing of multispectral 
Lidar (MSL) data associated with surface reflection from targeted surfaces composed of several known 
materials.
The problem addressed is the estimation of the positions and area distribution of each material.
In the Bayesian framework, appropriate prior distributions 
are assigned to the unknown model parameters and a Markov chain Monte Carlo method 
is used to sample the resulting posterior distribution.
The performance of the proposed algorithm is evaluated using synthetic
% and real
MSL signals,
for which single and multi-layered models are derived.
To evaluate the expected estimation performance associated with MSL signal analysis,
a Cramer-Rao lower bound associated with model considered is also derived, and compared with the experimental data.
Both the theoretical lower bound and the experimental analysis will be of primary assistance in future instrument design.
\end{abstract}


\section{Introduction}
Laser altimetry (or Lidar) is an acknowledged tool for extracting spatial structures from three-dimensional (3D) scenes, including forest canopies \cite{Mallet2009,Leeuwen2010}.
Using time-of-flight to create a distance profile, signal analysis can recover tree and canopy heights, leaf area indices (LAIs) and ground slope by analyzing the reflected photons from a target.
Conversely, passive multispectral (MSI) (dozen of wavelengths) and hyperspectral images (HSI) (hundreds of wavelengths) are widely used to extract spectral information about the scene which, for forest monitoring, can also provide useful parameters about the canopy composition and health (tree species, leaf chlorophyll content, water content, stress, among others) \cite{Lemaire2008,Hernandez2011}.
The most natural evolution to extract spatial and spectral information from sensed scenes is to couple Lidar data and multi/hyperspectral images \cite{Dalponte2008,Buckley2013}.
Although the fusion of Lidar data and HSIs can improve scene characterization, there are problems of data synchronization in space (alignment, resolution) and time (dynamic scene, change of observation conditions, etc).
For these reasons, multispectral Lidar (MSL) has recently received attention from the remote sensing community for its ability to extract both structural and spectral information
from 3D scenes \cite{Wallace2012, Hakala2012}.
 
The key advantage of MSL is the ability to provide information on the vertical
distribution of spectra, used to infer physiological processes directly linked to actual carbon
sequestration as well as carbon stocks~\cite{Wallace2014}. 
For example, a key capability is to directly measure and classify ground-based shrub infestation, which is difficult with conventional HSI as the lower spectral response is obscured by the 'first' signals returned from the top of the canopy. Preliminary trials with existing commercial LiDAR systems with three operational wavelengths have already taken place~\cite{Fleming2015}, but as we shall demonstrate here, such systems cannot yet provide all the necessary spectral and structural discrimination to directly measure these processes.

Another motivation for MSL is that HSI, even when fully synchronised, can only integrate the spectral response along the path of each optical ray, not measure the spectral response as a function of distance, e.g. depth into a forest canopy.
Multiple scattering effects cannot be neglected in some scenes with relief or containing multi-layered objects (such as trees), which complicates surface detection and quantification.
Most existing spectral unmixing (SU) techniques rely on a linear mixture
assumption \cite{Craig1994,Heinz2001,Eches2010a,Miao2007a,Yang2011} to identify the components (so-called endmembers) of an image and their proportions (abundances).
However, it has been shown that the classical linear mixing model (LMM) can be inaccurate when multiple scattering effects occur \cite{Keshava2002,Bioucas2012}.
Although polynomial models \cite{Altmann2012a,Altmann2014a} (including bilinear
models\cite{Somers2009,Nascimento2009,Fan2009,Halimi2010}) can be adapted for ``long range'' multiple scattering (in contrast to intimate mixtures models \cite{Hapke1981}), motivating the additional parameter constraints and designing accurate nonlinear unmixing methods are still challenging problems \cite{Dobigeon2014}.
Since MSL data present an additional dimension when compared to HSIs, we can expect a reduction of the SU problem complexity and thus an improvement in the target characterization performance.  

In \cite{Hernandez2007}, the authors proposed a Bayesian algorithm to estimate the positions and amplitudes in Lidar signals associated with a multi-layer target.
This method has been extended in \cite{Ramirez2012,Wallace2014} to MSL by first estimating the positions of the peaks (\emph{i.e.}, the layers), which were assumed to be the same in all spectral bands, then estimating the amplitudes of the peaks for each wavelength, and finally relating the estimated amplitude peaks to areas associated with each material that make up the target based on a linear mixing model.
The method has been applied to real MSL signals (four wavelengths) to analyze a conifer structure and has been shown to aid the recovery of bark and needle areas of the tree assuming it is modeled as a set of irregularly spaced layers. 

In this paper, we propose to investigate a SU problem applied to single- and multi-layered targets to estimate 
the positions and proportions of each (known) material. This problem extends the supervised SU problem of HSIs by 
also estimating the target position. 
Estimating the spectral variation of material signatures (unsupervised SU problem) is also 
of interest. For example, the estimation of physiological content, such as Chlorophyll concentration,
in forest canopies is of key concern, and this results in variable spectra for leaves and needles.
Extracting these spectra from MSL signals is a more challenging problem (probably more difficult than in HSIs due to the statistical properties of the noise and observation model) which is out of scope of this paper, but will require particular attention in future work.

In contrast with the classical additive Gaussian noise assumption used for HSIs, a Poisson noise model is 
more appropriate for MSL signals. Indeed, Lidar and thus MSL systems usually record, for each pixel/region of 
the scene, a histogram of time delays between emitted laser pulses and the detected photon
arrivals. Within each histogram bin, the number of detected photons follows a discrete distribution which can be approximated by a Poisson distribution due to the particle nature of light.
Using a Bayesian approach, appropriate prior distributions are chosen for the unknown parameters of the model 
considered here, \emph{i.e.}, the material areas, the surface positions and the background parameters.
The joint posterior distribution of these parameters is then derived. Since the classical Bayesian estimators 
cannot be easily computed from this joint posterior (mainly due to the model complexity), a Markov chain Monte 
Carlo (MCMC) method is used to generate samples according to the posterior of interest. 

More precisely, following the principles of the Gibbs sampler, samples are generated
according to the conditional distributions of the posterior.
Due to the possibly high correlations between the material proportions/areas, we propose
to use a Hamiltonian Monte Carlo (HMC) \cite{Duane1987} method to
sample according to some of the conditional distributions.
HMCs are powerful simulation strategies based on Hamiltonian dynamics which can
improve the convergence and mixing properties of classical MCMC methods (such as the
Gibbs sampler and the Metropolis-Hastings algorithm) \cite{Brooks2011,Robert2004}.
These methods have received growing interest in many
applications, especially when the number of parameters to be
estimated is large \cite{Neal1996,schmidt09funcfact}.
Classical HMC can only be used for unconstrained variables.
However, new HMC methods have been recently proposed to handle constrained
variables \cite[Chap. 5]{Brooks2011}, \cite{Hartmann2005,Brubaker2012} which allow HMCs to sample
according to the posterior of the Bayesian model proposed for
SU.
Finally, as in any MCMC method, the generated samples
are used to compute minimum mean square error (MMSE) estimators as well as measures
of uncertainties such as confidence intervals.

Predicting the parameter estimation performance is of prime interest for designing 
an estimation procedure but also assists with the instrument design.
Indeed, since MSL is a recent modality, it is important to identify the key parameters 
that have an influence on the material estimation performance (such as the necessary 
signal to background levels, the number of bands to be considered and the associated wavelengths).
To guide future instrument design, we consider a Cramer-Rao lower bound (CRLB) associated with 
the observation model and show that it can be used to approximate the estimation errors of the proposed Bayesian algorithm.

The remainder of the paper is organized as follows.
Section \ref{sec:model} introduces the observation model associated with MSL returns for a single-layered object to be analyzed.
Section \ref{sec:bayesian} presents the hierarchical Bayesian model associated with the proposed model and its posterior distribution.
The hybrid MCMC method used to sample from the posterior of interest is detailed in Section
\ref{sec:Gibbs}.
Section \ref{sec:CRLB} investigates the expected SU performance using Cramer-Rao lower bounds. 
Experimental results are shown and discussed in Section \ref{sec:experiments}. 
Conclusions are reported in Section \ref{sec:conclusion}.

\section{Problem formulation}
\label{sec:model} 
This section introduces the observation statistical model associated with MSL returns for a single-layered object which will be used in Sections \ref{sec:bayesian} and \ref{sec:Gibbs} to solve the SU problem of MSL data. Precisely, we consider a set of $L$ observed Lidar waveforms 
$\Vpix{\ell} = [\pix{\ell}{1},\ldots,\pix{\ell}{\nbbin}]\transp, \ell \in \left
\lbrace 1,\ldots,L \right \rbrace$ where $\nbbin$ is the number of temporal
(corresponding to range) bins.
To be precise, $\pix{\ell}{\nobin}$ is the photon count within the $\nobin$th bin of the $\ell$th spectral band considered. Let $\nobin_0$ be the position of a complex object surface or surfaces at a given range from the sensor, composed of $R$ materials whose spectral signatures (observed at $L$ wavelengths) are denoted as $\Vmat{r}=[\mat{1}{r},\ldots,\mat{\ell}{r}]\transp, r \in \{1,\ldots,R\}$. According to \cite{Hernandez2007}, each photon count $\pix{\ell}{\nobin}$ is assumed to be drawn from the following Poisson distribution 

\begin{eqnarray}
\label{eq:model0}
\pix{\ell}{\nobin} \sim \mathcal{P}\left(\sum_{r=1}^{R} w_{r}\mat{\ell}{r}g_{0,\ell}(\nobin) + b_{\ell}\right)
\end{eqnarray}

where $g_{0,\ell}(\cdot)$ is the photon impulse response whose shape can differ between wavelength channels, $w_{r}$ denotes the area of the $r$th material composing the object (relative to a reference area) and $b_{\ell}$ is a background and dark photon level, which is constant in all bins at a given wavelength.
Without loss of generality the photon impulse responses are assumed to be the same and modeled by the following piece-wise exponential
\begin{eqnarray}
\label{eq:impulse_resp0}
\begin{small}
g_{0,\ell}(\nobin)=\beta  \left\{
    \begin{array}{l}
        \exp^{-\frac{T_1^2}{2\sigma^2}}\exp^{\frac{\nobin+T_1-t_0}{\tau_1}}\quad \mbox{if } \nobin-t_0< -T_1 \\
				\exp^{-\frac{(\nobin-t_0)^2}{2\sigma^2}} \quad \mbox{if }   T_1 \leq \nobin-t_0< T_2 \\
				\exp^{-\frac{T_2^2}{2\sigma^2}}\exp^{\frac{\nobin-T_2-t_0}{\tau_2}} \quad \mbox{if } T_2 \leq \nobin-t_0< T_3 \\
				\exp^{-\frac{T_2^2}{2\sigma^2}}\exp^{\frac{T_2-T_3}{\tau_2}}\exp^{\frac{\nobin-T_3-t_0}{\tau_3}} \quad \mbox{else }\\
    \end{array}
\right.\nonumber
\end{small}
\end{eqnarray}
where $\bphi=[T_1,T_2,T_3,\tau_1,\tau_2,\tau_3,\sigma^2,\beta]\transp$ is a positive hyperparameter vector. 
In this study, we assume that the shape parameters of the model (which is appropriate for our own lidar sensors) are fixed and known from the instrumental response.
As seen later in Fig. \ref{fig:impulse_approx} the shape is slightly asymmetrical when compared to the more common Gaussian model \cite{Wagner2006} which is used for higher power, longer pulse duration lidar systems. The consideration of a piece-wise exponential approximation is motivated by the actual shape of our instrumental impulse response but it is worth mentioning that any positive analytical approximation, even different approximations for the different wavelengths, could be used instead without modifying the proposed Bayesian estimation procedure. The consideration of a single impulse response model is done only for ease of reading.
While a Gaussian approximation can lead to a slight bias in depth estimation, we shall use a Gaussian
approximation in Section \ref{sec:CRLB} to compute the Cramer-Rao bounds since the Gaussian approximation is twice differentiable (in contrast to the piece-wise exponential approximation), which simplifies the derivation of the bounds. 

Due to physical considerations, the relative areas are assumed to satisfy the following positivity constraints 
\begin{eqnarray}
\label{eq:pos_const}
w_{r} \geq 0, \quad r=1,\ldots,R.
\end{eqnarray}
Similarly, the average background level in each channel satisfies $b_{\ell}\geq 0, \ell=1,\ldots,\nbband$. Spectral unmixing of the MSL data consists of estimating the position (\emph{i.e.}, $t_0$) of the target, the area $w_{r}$ of each component $\Vmat{r}$ of the target, as well as the background levels from the observed data in $\MATpix=[\Vpix{1},\ldots,\Vpix{\nbband}]$. The next section studies a Bayesian model to estimate the unknown parameters in \eqref{eq:model0} while ensuring the positivity constraints mentioned above.

%%%%%%%%%%%%%%%%%%%%%%%%%%%%%%%%%%%%%%%%%%%%%%%%%%%%%%%%%%%%%%%%%%%%%
%%%%%%%%%%%%%%%%%%%%%%%%%%%%%%%%%%%%%%%%%%%%%%%%%%%%%%%%%%%%%%%%%%%%%
\section{Bayesian model}
\label{sec:bayesian}
The unknown parameter vector associated with \eqref{eq:model0} contains the relative areas $\bfw=[w_{1},\ldots,w_{R}]\transp$, the position of the target $t_0$ and  background levels $\bfb=[b_{1},\ldots,b_{\nbband}]\transp$ (satisfying positivity constraints). This section summarizes the likelihood and the parameter priors associated with these parameters. 

\subsection{Likelihood}
\label{subsec:likelihood}
Assuming that the detected photon counts/noise realizations, conditioned on their mean in all bins and for all wavelengths, are independent, Eq. \eqref{eq:model0} leads to 
\begin{eqnarray}
\label{eq:likelihood}
P(\MATpix|\MATmat,\bfw,\bfb,t_0) = \prod_{\ell=1}^{\nbband} \prod_{\nobin=1}^{\nbbin} \dfrac{\lambda_{\ell,\nobin}^{\pix{\ell}{\nobin}}}{\pix{\ell}{\nobin}!} \exp^{-\lambda_{\ell,\nobin}}
\end{eqnarray}
where $\MATmat=[\Vmat{1},\ldots,\Vmat{R}]$, $\lambda_{\ell,t}=\Vmat{\ell,:}\bfw g_{0,\ell}(\nobin) + b_{\ell}$ and $\Vmat{\ell,:}$ denotes the $\ell$th row of $\MATmat$.

\subsection{Prior for the target position}
Since we don't have prior information about the position of the target, the following uniform distribution 
\begin{eqnarray}
\label{eq:prior_t}
t_0 \sim \mathcal{U}_{(1;\nbbin)}(t_0)
\end{eqnarray}
is assigned to $t_0$. Note that the position $t_0$ is a real variable that is not restricted to the integer values in $(1;\nbbin)$.
\subsection{Prior for the relative areas}
To reflect the lack of prior knowledge about the areas, the following truncated Gaussian prior
\begin{eqnarray}
w_{r}|\alpha^2 \sim \mathcal{N}_{\bbR^+}(w_{r};0,\alpha^2)
\end{eqnarray}  
is assigned to each area $w_{r}$, where $\mathcal{N}_{\bbR^+}(\cdot;0,\alpha^2)$ denotes the Gaussian distribution restricted to $\bbR^+$, which hidden mean and variance $0$ and $\alpha^2$, respectively. The hyperparameter $\alpha^2$ is shared by all the parameters $w_{r}$ and is arbitrarily fixed to a large value to ensure a weakly informative prior.
Assuming prior independence for the unknown areas yields
\begin{eqnarray}
\label{eq:joint_prior_w}
f(\bfw|\alpha^2) \propto \left(\dfrac{1}{\alpha^2}\right)^{\frac{R}{2}}\exp^{-\frac{1}{2\alpha^2}\bfw\transp\bfw}\Indicfun{\left(\bbR^+\right)^R}{\bfw}
\end{eqnarray}  
where $\propto$ means ''proportional to'' and $\Indicfun{\left(\bbR^+\right)^R}{\cdot}$ is the indicator function defined on $\left(\bbR^+\right)^R$.
\subsection{Prior for the background level}
Similarly, assigning a truncated Gaussian prior to each background level and assuming prior independence between these parameters leads to 
\begin{eqnarray}
\label{eq:joint_prior_b}
f(\bfb|\gamma^2) \propto \left(\dfrac{1}{\gamma^2}\right)^{\frac{\nbband}{2}}\exp^{-\frac{1}{2\gamma^2}\bfb\transp\bfb}\Indicfun{\left(\bbR^+\right)^\nbband}{\bfb}
\end{eqnarray}
where the hyperparameter $\gamma^2$ is shared by all the parameters $b_{\ell}$ and is arbitrarily fixed to a large value to ensure a weakly informative prior.

In this paper, we assume that prior knowledge about the relative areas and background levels is limited. We use weakly informative Gaussian priors restricted to the positive orthant to reflect this lack of knowledge and reduce the estimation bias while ensuring the positivity constraints are satisfied. However any proper weakly informative prior distributions could have been used, leading to similar estimation results. Similarly, if useful information about the unknown parameters is available, more informative (possibly physics-based) priors can be used instead of \eqref{eq:prior_t} and \eqref{eq:joint_prior_w} (target position and composition) and \eqref{eq:joint_prior_b} (background levels).

\subsection{Joint posterior distribution}

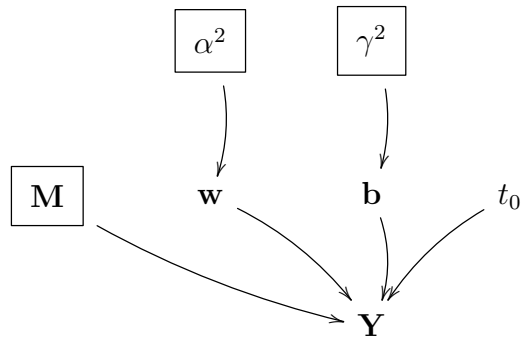
\begin{figure}[h!]
\centerline{ \xymatrix{
  & *+<0.05in>+[F-]+{\alpha^2} \ar@/^/[d] & *+<0.05in>+[F-]+{\gamma^2} \ar@/^/[d] &\\
  *+<0.05in>+[F-]+{\MATmat} \ar@/_/[rrd]   &   \bfw \ar@/^/[rd]& \bfb \ar@/^/[d] & t_0 \ar@/_/[ld] \\
  & & \MATpix   & }
} \caption{DAG for the parameter and hyperparameter priors (the
fixed parameters appear in boxes).} \label{fig:DAG}
\end{figure}

The resulting directed acyclic graph (DAG) associated with the
proposed Bayesian model is depicted in Fig. \ref{fig:DAG}. The joint posterior distribution of the unknown parameter vector $\paramvect=\left\lbrace  \bfw,\bfb, t_0\right\rbrace$ can be computed using
\begin{eqnarray}
\label{eq:posterior}
f(\paramvect|\MATpix,\MATmat,\alpha,\beta) \propto P(\MATpix|\MATmat,\bfw,\bfb,t_0)f(\paramvect|\alpha,\beta)
\end{eqnarray}
where $P(\MATpix|\MATmat,\bfw,\bfb,t_0)$ has been defined in \eqref{eq:likelihood} and $f(\paramvect|\alpha,\beta)=f(\bfw|\alpha^2)f(\bfb|\beta^2)f(t_0)$.

Unfortunately, it is difficult to obtain closed form expressions for
standard Bayesian estimators associated with \eqref{eq:posterior}.
In this paper, we propose to use efficient Markov Chain Monte Carlo
(MCMC) methods to generate samples asymptotically distributed
according to \eqref{eq:posterior}, employing a Gibbs sampler.
The principle of the Gibbs sampler is to
sample according to the conditional distributions of the posterior
of interest \cite[Chap. 10]{Robert2004}.
Due to the high correlation between the elements of $\bfw$ (which will be discussed later in the paper), 
we use an Hamiltonian Monte Carlo (HMC) method which improves the mixing properties of the sampling 
procedure (compared to a classical Metropolis-within-Gibbs sampler using 
Gaussian random walks). This method is detailed in the next section.
%%%%%%%%%%%%%%%%%%%%%%%%%%%%%%%%%%%%%%%%%%%%%%%%%%%%%%%%%%%%%%%%%%%%%
%%%%%%%%%%%%%%%%%%%%%%%%%%%%%%%%%%%%%%%%%%%%%%%%%%%%%%%%%%%%%%%%%%%%%

%%%%%%%%%%%%%%%%%%%%%%%%%%%%%%%%%%%%%%%%%%%%%%%%%%%%%%%%%%%%%%%%%%%%%
%%%%%%%%%%%%%%%%%%%%%%%%%%%%%%%%%%%%%%%%%%%%%%%%%%%%%%%%%%%%%%%%%%%%%
\section{Bayesian inference using a Constrained Hamiltonian Monte Carlo method}
\label{sec:Gibbs}
In this Section, we develop an HMC-within Gibbs sampler to generate samples according to \eqref{eq:posterior} and 
estimate the unknown parameters involved in \eqref{eq:model0} in order to solve the SU problem 
of MSL data for a single-layered object.
The proposed sampler consists of three steps to update sequentially $\bfw,t_0$ 
and $\bfb$ and is summarised in Algo. \ref{algo:bayesian}. 

\subsection{Sampling the areas}
The full conditional distribution of $\bfw$ is given by 
\begin{eqnarray}
\label{eq:joint_post_area}
f(\bfw|\MATpix,\MATmat,t_0,\bfb,\alpha^2) \propto  P(\MATpix|\MATmat,\bfw,\bfb,t_0)f(\bfw|\alpha^2)
\end{eqnarray}
and is not a standard distribution which is easy to sample.
Consequently, it is the norm to use an accept/reject procedure to update $\bfb$.
A classical and simple approach would use a multivariate Gaussian random walk.
However, in practice the elements of $\bfw$ can be highly correlated, especially when the materials present similar spectral signatures (which will be the case for vegetation targets).
Consequently, a Hamiltonian Monte Carlo method \cite[Chap. 5]{Brooks2011} is preferred to improve the mixing properties of the sampler.
The principle of HMCs is to introduce auxiliary, or momentum, variables, and perform a Metropolis-Hastings move in a higher dimensional parameter space.
The proposal distribution is then built to take into account the shape of the target distribution \eqref{eq:joint_post_area}.
Due to the area constraints \eqref{eq:pos_const}, a constrained HMC must be used.
In this paper, we used a constrained HMC (CHMC) scheme similar to that described in \cite{Altmann2014a} and thus consider the following potential energy function 
\begin{eqnarray}
U(\bfw)= -\sum_{\ell,\nobin} \pix{\ell}{\nobin}\log(\lambda_{\ell,\nobin}) + \lambda_{\ell,\nobin} +\dfrac{\bfw\transp\bfw}{2\alpha^2}
\end{eqnarray} 
to simulate Hamiltonian dynamics and compute the appropriate acceptance ratio (see \cite{Altmann2014a,Brooks2011} for technical details).
This choice of potential energy function ensures that $U(\bfw)=-\log\left(f(\bfw|\MATpix,\MATmat,t_0,\bfb,\alpha^2)\right) + c$ where $c$ is a positive constant. 
Note that the proposed CHMC step can be applied for any differentiable relative area prior. Metropolis-Hastings or more complex HMC moves should be used instead when considering non-differentiable priors instead of \eqref{eq:joint_prior_w}. 

\subsection{Sampling the target position}
It can be shown from \eqref{eq:posterior} that 
\begin{eqnarray}
\label{eq:post_t0}
f(t_0|\MATpix,\MATmat,\bfw,\bfb,\alpha^2) \propto  P(\MATpix|\MATmat,\bfw,\bfb,t_0)f(t_0)
\end{eqnarray}
is not a standard distribution and an accept/reject procedure must be used to update the target position $t_0$. 
We use a Gaussian random walk to update this parameter.
More precisely, a doubly truncated Gaussian proposal is considered to ensure each candidate belongs to the admissible set $(1;\nbbin)$ and the variance of the proposal is adjusted during the
burn-in period of the sampler to obtain an acceptance rate close to $0.45$, as recommended
in \cite[p. 8]{Robertmcmc}.

\subsection{Sampling the background levels}
It can be shown from \eqref{eq:posterior} that
\begin{eqnarray}
\label{eq:joint_post_b}
f(\bfb|\MATpix,\MATmat,\bfw,t_0,\alpha^2) = \prod_{\ell=1}^{L} f(b_{\ell}|\Vpix{\ell},\Vmat{\ell,:},\bfw,t_0,\alpha^2),
\end{eqnarray}
where\\ $f(b_{\ell}|\Vpix{\ell},\Vmat{\ell,:},\bfw,t_0,\alpha^2)$
\begin{eqnarray}
\label{eq:post_bl}
\propto \prod_{\nobin=1}^{\nbbin} \lambda_{\ell,\nobin}^{\pix{\ell}{\nobin}}\exp^{-\lambda_{\ell,\nobin}}\exp^{-\frac{ b_{\ell}^2}{2\gamma^2}}\Indicfun{\bbR^+}{b_{\ell}}
\end{eqnarray}

Sampling from \eqref{eq:post_bl} is again not straightforward and Gaussian random walks are used to update the background levels, similar to the position $t_0$.  
However the background levels are a posteriori independent and can be updated in parallel.
Similar to the target position update, the variances of the $L$ parallel Gaussian random walk
procedures are set during the burn-in period of the sampler to obtain an acceptance rate close to $0.45$.

\begin{algogo}{Gibbs Sampling Algorithm (single layer)}
     \label{algo:bayesian}
     \begin{algorithmic}[1]
        \STATE \underline{Fixed input parameters:} $\MATmat,\bphi,\alpha^2,\gamma^2$, 
				number of burn-in iterations $N_{\textrm{bi}}$, total number of iterations$N_{\textrm{MC}}$
				\STATE \underline{Initialization ($i=0$)}
        \begin{itemize}
        \item Set $\bfw^{(0)},t_0^{(0)},\bfb^{(0)}$
        \end{itemize}
        \STATE \underline{Iterations ($1 \leq i \leq N_{\textrm{MC}}$)}
        \STATE Sample $\bfw^{(i)}$ from the pdf in \eqref{eq:joint_post_area} and CHMC
        \STATE Sample $t_0^{(i)}$ from the pdf in \eqref{eq:post_t0} and a Gaussian random walk
        \STATE Sample $\bfb^{(i)}$ from the pdfs in
            \eqref{eq:post_bl} and Gaussian random walks
        \STATE Set $i = i+1$.
\end{algorithmic}
\end{algogo}

After generating $N_{\textrm{MC}}$ samples using the procedures
detailed above and removing $N_{\textrm{bi}}$ iterations associated
with the burn-in period of the sampler ($N_{\textrm{bi}}$ has been
set from preliminary runs), the MMSE estimator of the unknown parameters can be
approximated by computing the empirical averages of the remaining 
samples. The minimal length of the burn-in period can be determined using several convergence diagnostics \cite{Robertmcmc} but the number of initial samples to be discarded generally varies between data sets and can be thus difficult to adjust in advance without overestimating it.
In previous work on parallel acceleration of RJMCMC algorithms we have investigated the application of such diagnostics to our monochromatic LiDAR signals \cite{Ye2013}, but as absolute speed of processing is not a major concern here, the length of the burn-in period is assessed visually from the preliminary runs and fixed to ensure that the sampler has converged. The total number of iterations has then been set to obtain accurate approximations (computed over $4000$ samples in Section \ref{sec:experiments}) of the MMSE estimators.

\section{Predicting unmixing performance}
\label{sec:CRLB}
This section studies a Cramer-Rao lower bound associated with the observation model 
\eqref{eq:model0} which can be used to assess the performance of methods that aim to solve 
the SU problem of MSL data considered in this paper as well as to assist with the future instrument design. Precisely, Section \ref{subsec:CRLB_def} recalls the definition of the CRLB for the problem of interest and Section \ref{subsec:perf_analysis} discusses the impact of several key parameters on the expected parameter estimation performance of SU methods.

\subsection{Cramer-Rao lower bound}
\label{subsec:CRLB_def} 
Prediction of the unmixing performance is necessary to assist in the design of
a lidar system for a specific application, identifying those parameters 
that have the most significant impact.
In deriving a CRLB, we propose firstly to relax the impulse response model in \eqref{eq:impulse_resp0} by considering the following Gaussian approximation 
\begin{eqnarray}
\label{eq:impulse_resp1}
\tilde{g}_{0,\ell}(\nobin)=\beta  
				\exp^{-\frac{(\nobin-t_0)^2}{2\sigma^2}} \approx g_{0,\ell}(\nobin) \quad  \forall \nobin \in (1;\nbbin),\forall \ell.
\end{eqnarray}
This simplifies the CRLB derivation and does not significantly bias the prediction as the piece-wise exponential impulse response can be accurately approximated by a Gaussian function.
The CRLB associated with any unbiased estimator $\hat{\paramvect}$ of the parameter vector $\paramvect$ involved in the mixing model \eqref{eq:model0} (having replaced $g_{0,\ell}(\nobin)$ by $\tilde{g}_{0,\ell}(\nobin)$) and constructed from $\MATpix$ is given by
 \begin{eqnarray}
\label{eq:CRLB}
\textrm{CRLB}(\paramvect) = \bfJ_{\textrm{F}}^{-1}
\end{eqnarray}
where $\bfJ_{\textrm{F}}$ is the Fisher information matrix whose elements are\footnote{The Fisher information matrix $\bfJ_{\textrm{F}}$ is derived in the Appendix}
\begin{eqnarray}
\left[\bfJ_{\textrm{F}} \right]_{i,j} = - \textrm{E}_{\MATpix|\paramvect} \left[ \dfrac{\partial^2 \log f(\MATpix|\paramvect)}{\partial\theta_i \partial \theta_j}\right], i,j =1,\ldots,R+\nbband+1.\nonumber
\end{eqnarray}
The $i$th diagonal element of the CRLB matrix in \eqref{eq:CRLB} provides a lower bound for the variance of $\hat{\theta_i}$, given that $\hat{\paramvect}$ is an unbiased estimator of $\paramvect$.
Of course, the Bayesian estimation procedure proposed in Sections \ref{sec:bayesian} and \ref{sec:Gibbs} does not provide a strictly unbiased estimator of $\paramvect$ and a Bayesian CRLB should have been considered instead of \eqref{eq:CRLB}. However, due to the weakly informative priors chosen in Section \ref{sec:bayesian}, when the actual vector $\paramvect$ is far enough from the boundaries of the admissible set (defined by the positivity constraints), the bias of the proposed estimation procedure can be neglected and the resulting estimator achieves the CRLB (as will be shown in Section \ref{sec:experiments}). Moreover, the CRLB in \eqref{eq:CRLB} can also provide information about the estimation performance of a potential optimization method that could be proposed to estimate $\paramvect$ based on maximum likelihood estimation (MLE).

\subsection{Performance analysis}
\label{subsec:perf_analysis}
\begin{figure}[h!]
  \centering
  \includegraphics[width=\columnwidth]{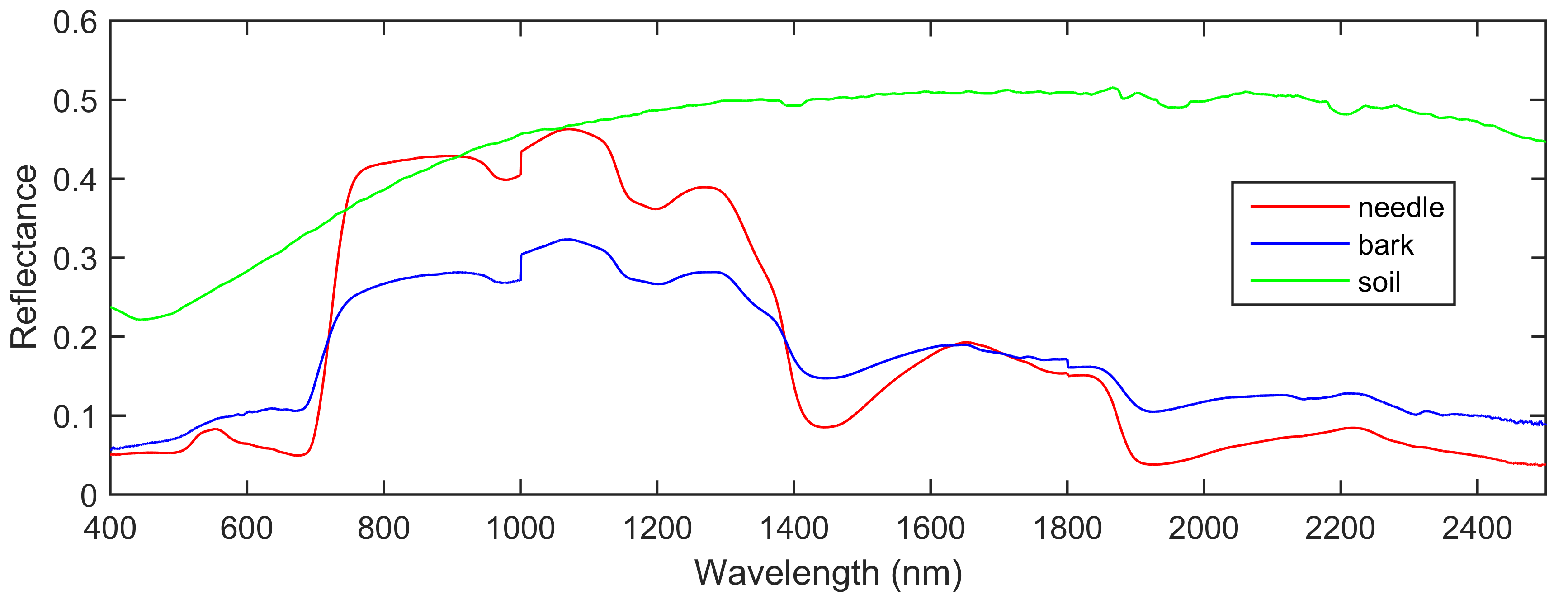}
  \caption{Spectral signatures of needle (red), bark (blue) and soil (green) considered in the synthetic data.}
  \label{fig:endmembers}
\end{figure}

\begin{figure}[h!]
  \centering
  \includegraphics[width=\columnwidth]{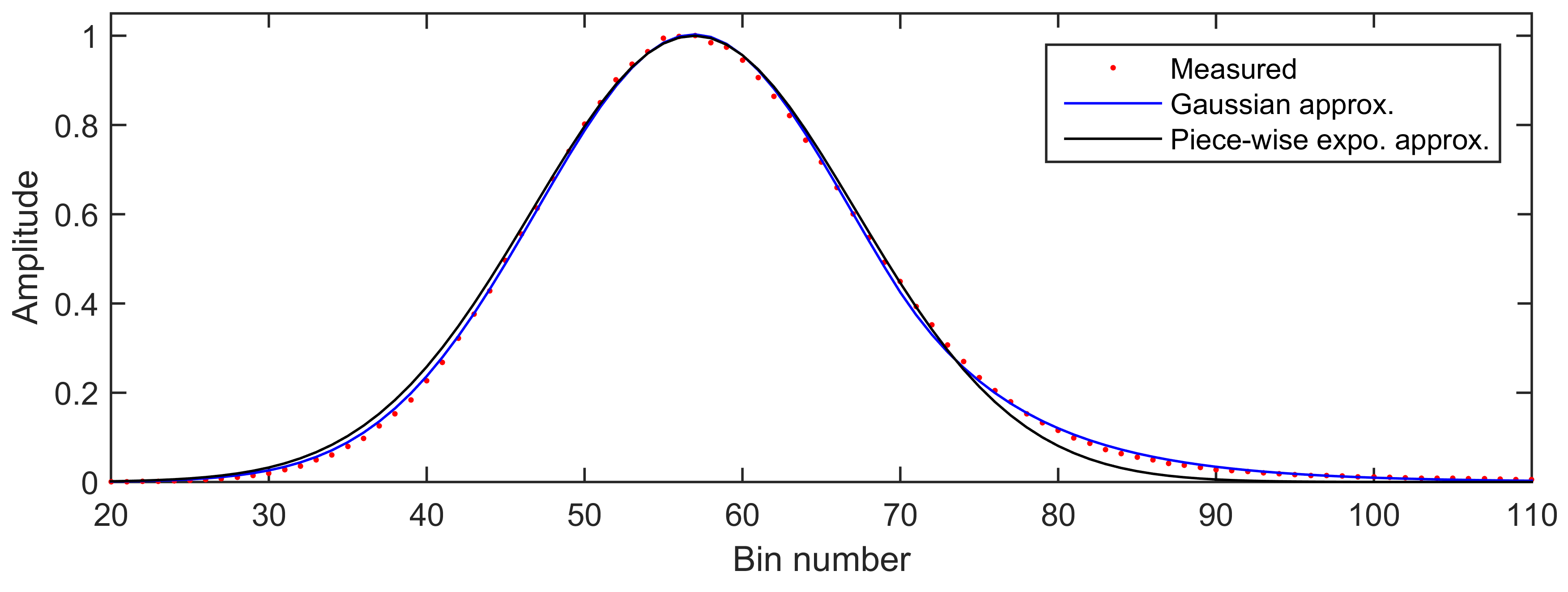}
  \caption{Real normalized photon impulse response (red dots) and approximations using piece-wise exponential (blue line) and Gaussian (black line) functions.}
  \label{fig:impulse_approx}
\end{figure}

\begin{figure}[h!]
  \centering
  \includegraphics[width=\columnwidth]{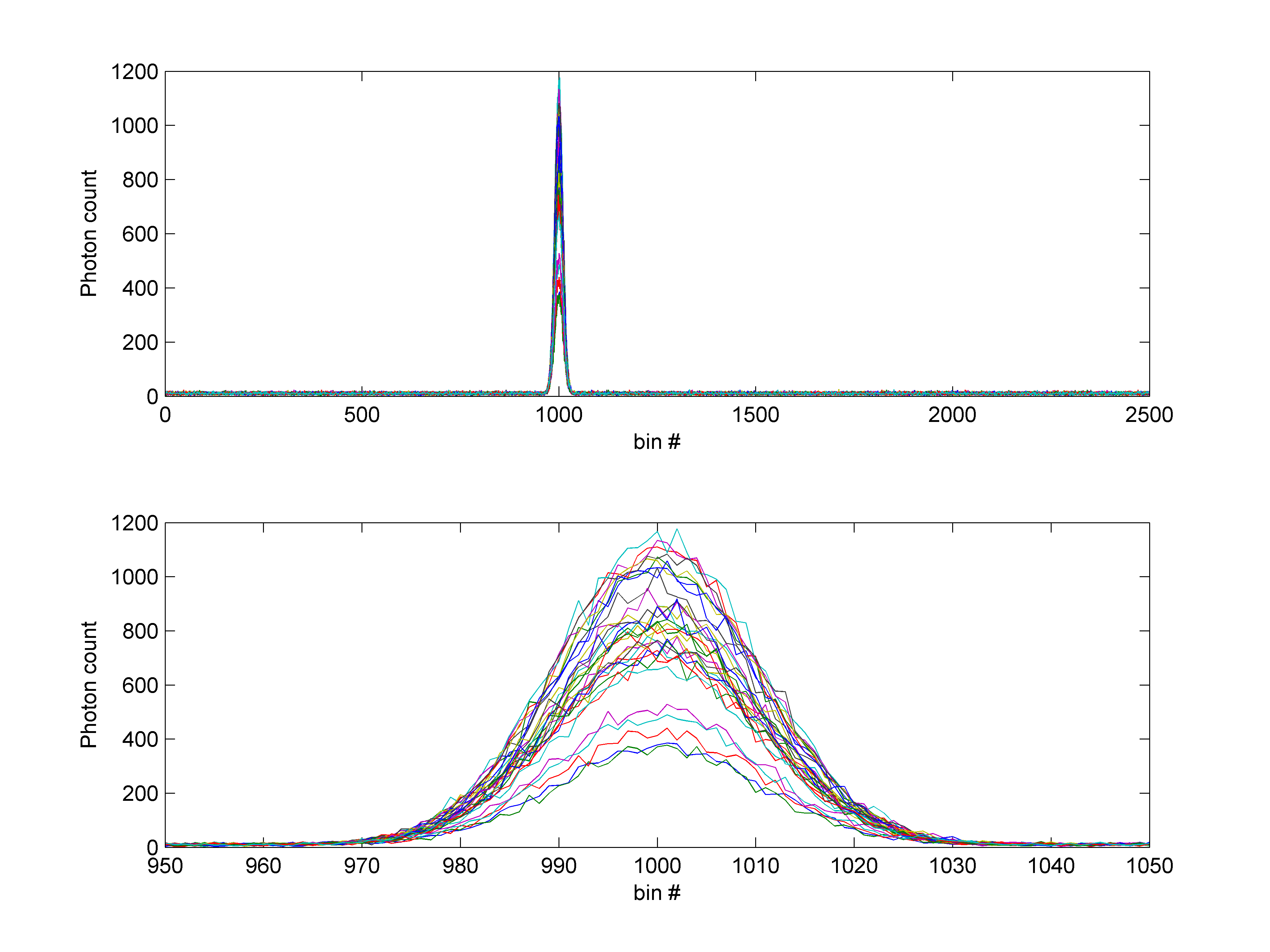}
  \caption{Example of MSL signals for $L=32$ spectral bands and a target located at $t_0=1000$ (the bottom subplot is a zoom around the actual target location). The different colors correspond to different spectral bands.}
  \label{fig:example_MSL}
\end{figure}

We first investigate the performance of spectral unmixing of MSL signals for a single layered, artificial target composed of $R=3$ materials, specifically needles, bark and soil.
These materials have been chosen because of our interests in forest canopy monitoring using MSL
signals \cite{Wallace2014}.
The reflectance spectra of these materials, observed at equally spaced spectral bands ranging from $400$nm to $2500$nm are depicted in Fig. \ref{fig:endmembers}. It is interesting to note the strong similarity between the bark and needles spectral signatures, which can make their discrimination difficult and highlights the need of multiple wavelengths to quantify these materials.
The maximum number of spectral bands considered in this paper has been set to $L_{\textrm{max}}=32$, which is a realistic value for a short-term real measurement campaign (the current instrument uses only $4$ wavelengths).
The relative area of needles (resp. bark and soil) has been arbitrarily set to $w_{n}=0.2$ (resp. $w_{b}=0.3$ and $w_{s}=0.4$). 
These relative areas do not necessarily sum to one as we consider that part of the laser light can penetrate through the target (semi-transparent target or a significant gap fraction)
and may not be further reflected (single hit assumption).
The fixed model parameters have been fixed from experiments to 
\begin{eqnarray}
\label{eq:param_exp}
\left\{
    \begin{array}{l}
				\nbbin=2500 \\
				\beta=3000 \\
				t_0=1000\\
				b_{\ell}=b=10, \quad \forall \ell\\
    \end{array}
\right.
\end{eqnarray}
These parameters will be fixed  in the remainder of the paper unless otherwise specified.
The impulse response parameters have been set to $T_1=402,T_2=12.5,T_3=239,\tau_1=395,\tau_2=7.9,\tau_3=1595$ and $\sigma^2=105.82$
for the piece-wise exponential approximation in \eqref{eq:impulse_resp0} and $\sigma^2=105.68$ for the Gaussian approximation in \eqref{eq:impulse_resp1}. These parameters have been obtained by fitting the experimental impulse response measured in \cite{Hernandez2007}. Fig. \ref{fig:impulse_approx} shows that the Gaussian approximation provides a good estimate of the experimental impulse response, although the piece-wise exponential better fits the experimental curve.  
Fig. \ref{fig:example_MSL} shows an example of MSL data generated using the parameters in \eqref{eq:param_exp} and the impulse response approximation in \eqref{eq:impulse_resp0}. 
Table \ref{tab:CRLB} shows the predicted estimation performance for the unknown parameters of interest (\emph{i.e.}, $\bfw=[w_n,w_b,w_s]\transp$ and $t_0$) assuming the Gaussian approximation of the photon impulse response \eqref{eq:impulse_resp1}. The relative errors provided in the bottom row of Table \ref{tab:CRLB} are computed by dividing the square root of the CRLBs by the actual values of the parameters. The relative estimation errors of the background levels (not presented here) are lower than $1\%$. 
\begin{table}[h!]
\renewcommand{\arraystretch}{1.2}
\begin{footnotesize}
\begin{center}
\caption{Estimation performance ($L=32$)
.\label{tab:CRLB}}
\begin{tabular}{|c|c|c|c|c|}
\cline{2-5}
\multicolumn{1}{c|}{} &  $w_n$  & $w_b$ & $w_s$ & $t_0$ \\
\hline
\multicolumn{1}{|c|}{Actual value} & $0.2$ & $0.3$ & $0.4$ & $1500$\\
\hline
\multicolumn{1}{|c|}{CRLB} & $2.6 \times 10^{-4}$& $0.001$ & $3.6 \times 10^{-5}$ &   $1.8 \times 10^{-4}$ \\
\hline
\multicolumn{1}{|c|}{Rel. err. (\%)}& 8.0 & 10.7 & 1.5  & $<10^{-5}$\\
\hline
\end{tabular}
\end{center}
\end{footnotesize}
\vspace{-0.4cm}
\end{table}

\begin{figure}[h!]
  \centering
  \includegraphics[width=\columnwidth]{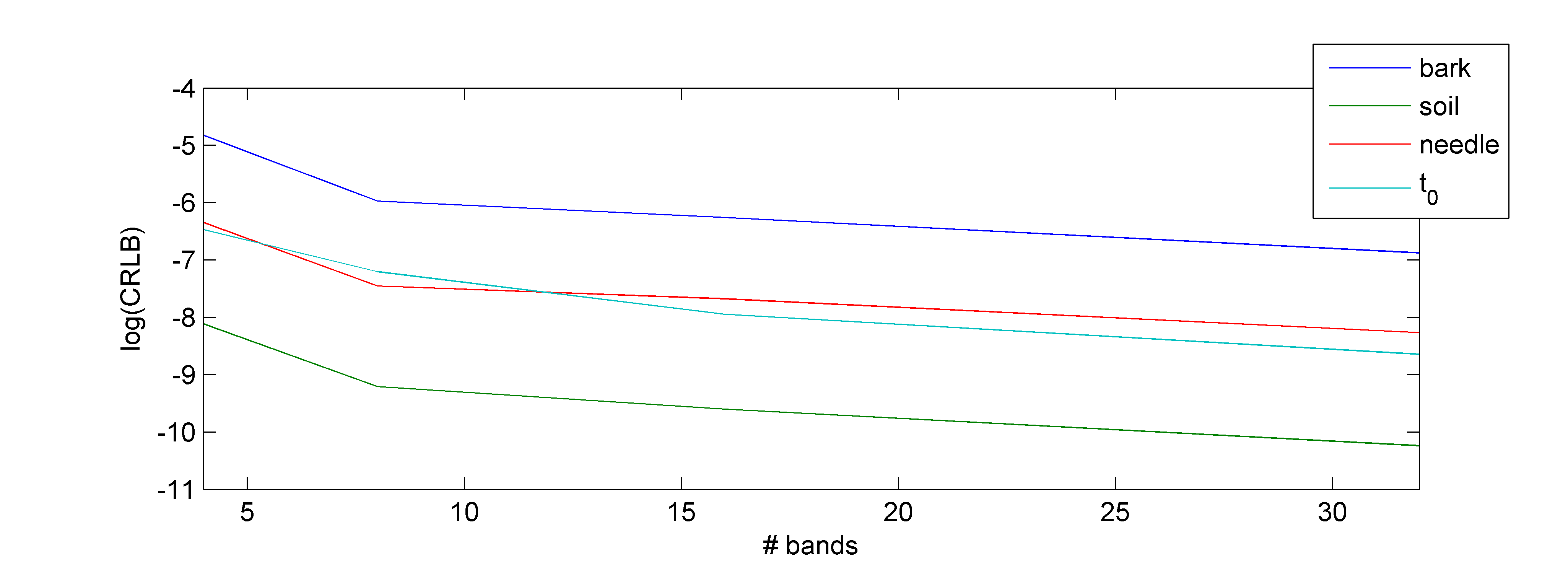}
  \caption{Evolution of the CRLB for $\bfw$ and $t_0$ as a function of the number of bands $L$ ($\beta=3000, b_{\ell}=10, \forall \ell$).}
  \label{fig:CRLB_L}
\end{figure}

\begin{figure}[h!]
  \centering
  \includegraphics[width=\columnwidth]{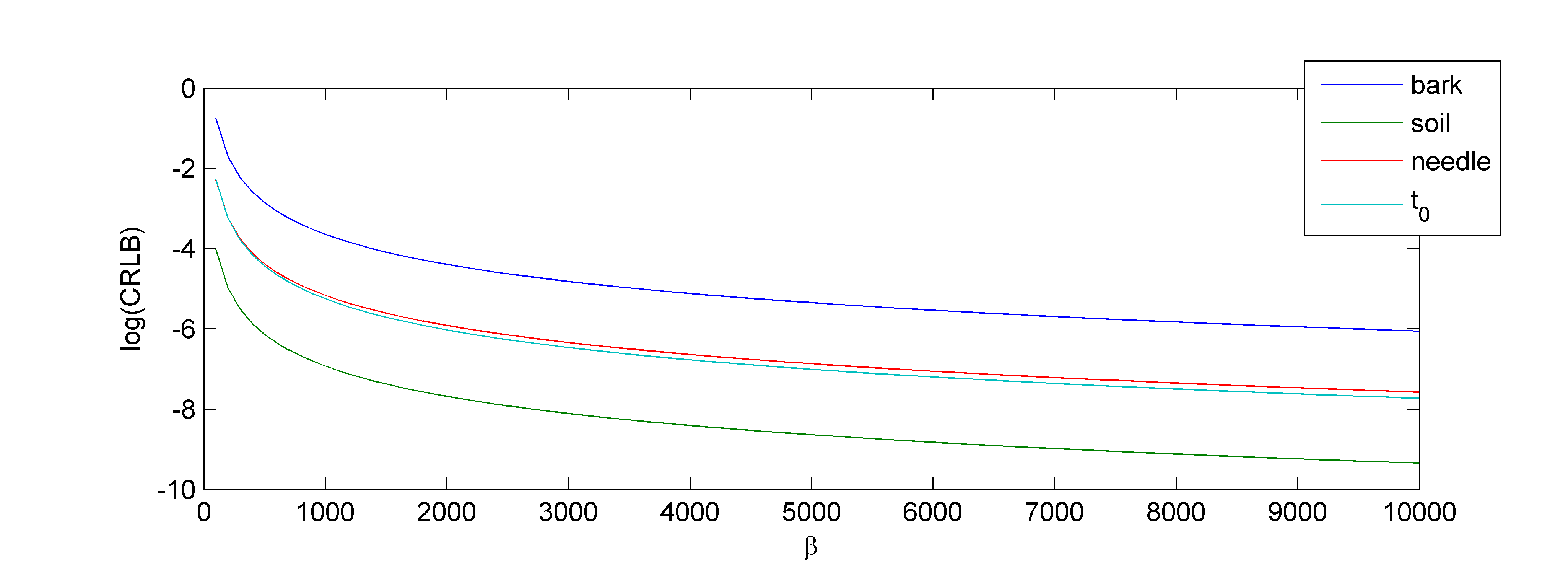}
  \caption{Evolution of the CRLB for $\bfw$ and $t_0$ as a function of $\beta$ ($L=32, b_{\ell}=10$).}
  \label{fig:CRLB_bet}
\end{figure}

\begin{figure}[h!]
  \centering
  \includegraphics[width=\columnwidth]{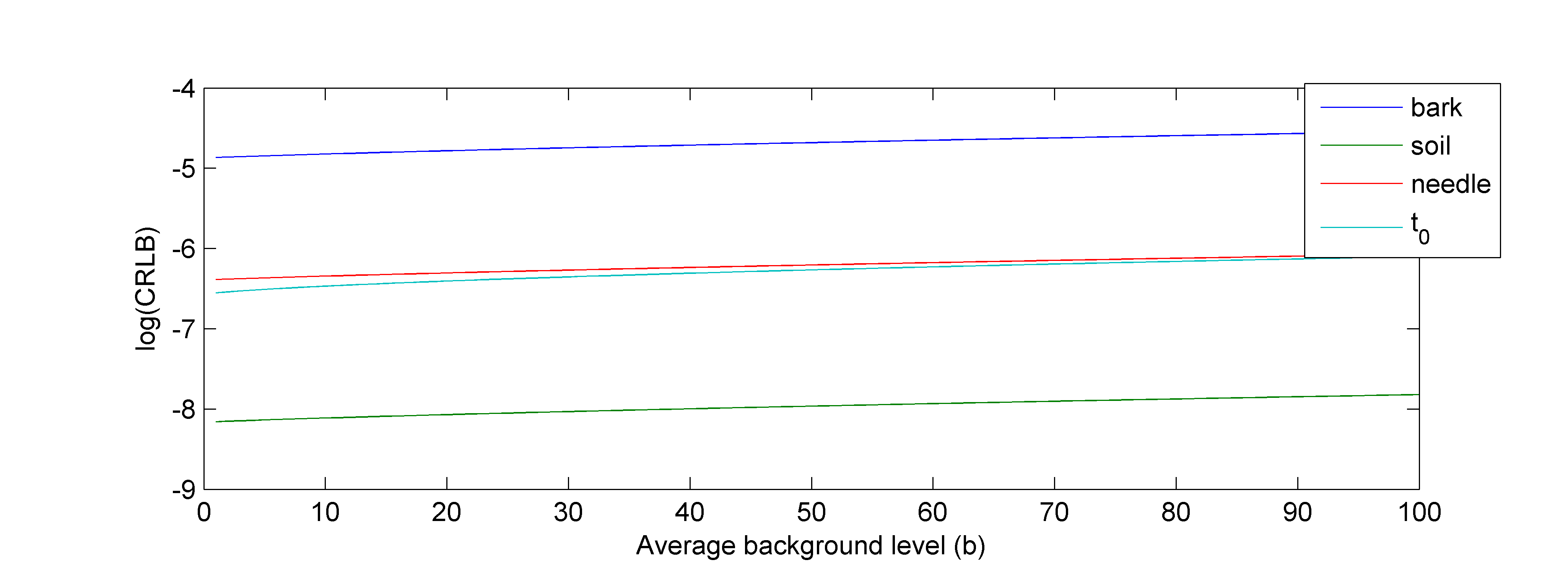}
  \caption{Evolution of the CRLB for $\bfw$ and $t_0$ as a function of $b=b_{\ell}, \forall \ell$ ($L=32, \beta=3000$).}
  \label{fig:CRLB_b}
\end{figure}

These results show that the estimation errors associated with the target position are usually much lower than those associated with the material areas.
Moreover, although the amplitude of the peak for each wavelength is much larger than the background level (see Fig. \ref{fig:example_MSL}), the CRLB predicts possibly large estimation errors, especially for the needle and bark areas.
This can be partially explained by the fact that the soil reflectance spectrum presents an average energy which is higher than those of needles and bark (and thus $w_s>w_b>w_n$) but also and mainly because of the high correlation between the bark and needle spectra, which complicates their quantification.
Figs. \ref{fig:CRLB_L} to \ref{fig:CRLB_b} show the predicted estimation errors for different values of the key parameters $\beta$ (which is related to the number of photons emitted by the laser sources), the number of spectral bands (equally spaced between $400$ nm and $2500$ nm) and the average background level $b$ (assumed to be the same in all spectral bands and bins).
Fig. \ref{fig:CRLB_L} shows that the estimation performance generally increases with the number of spectral bands.
This result is well known for SU of multispectral and hyperspectral images and is here demonstrated for MSL signals.
However, the performance improvement can vary, depending on the additional spectral bands considered. For instance, additional bands containing similar spectral information may not significantly improve the unmixing results (\emph{e.g.}, several wavelengths between $2000$ nm and $2500$ nm for the three materials in Fig. \ref{fig:endmembers}).
Fig. \ref{fig:CRLB_bet} shows that the parameter $\beta$ has a significant impact on the SU performance.
This parameter increases with the amplitude and the number of laser source pulses used to acquire the data and decreases with the average distance between the source and the target (as the probability of recording a reflected photon decreases).
Increasing $\beta$ leads to better area estimation but can require a longer target exposure (which can be problematic for airborne sensors for instance).
Finally, Fig. \ref{fig:CRLB_b} shows that for relatively small values of background levels (compared to $\beta$), the SU performance is not significantly degraded when the background increases.
This background level (which depends primarily on the background (e.g. solar) radiation as well as
the instrument design) is expected to be quite small in practice. 

\clearpage
\section{Experiments}
\label{sec:experiments}
\begin{figure}[h!]
  \centering
  \includegraphics[width=\columnwidth]{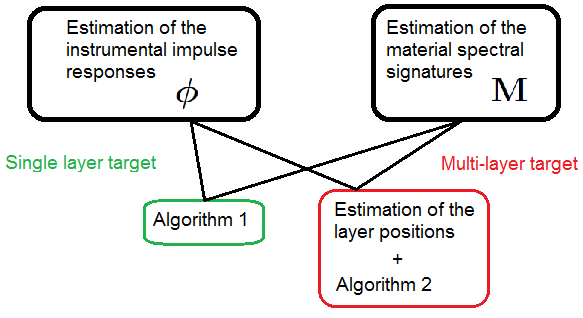}
  \caption{Main steps of the method proposed for spectral unmixing of MSL data associated with single and multi-layer targets.}
  \label{fig:algo_steps}
\end{figure}

In this section, we first apply (Section \ref{subsec:simu_single_layer}) the proposed SU method to synthetic MSL data associated with a single-layered target and compare its estimation performance to that predicted by the CLRB presented in Section \ref{sec:CRLB} and to that of a state-of-the art method \cite{Wallace2014}. In Section \ref{subsec:multilayer}, we investigate the case of a multi-layered target whose layer positions are assumed to be known. Although the current algorithm does not estimate the positions of multiple surfaces, Section \ref{subsec:multilayer} aims to illustrate how the proposed method can be extended to analyze more complex objects. The main steps of the methodology (proposed to analyze single and multi-layer target) are summarised in Fig. \ref{fig:algo_steps}.
\subsection{Single layer target}
\label{subsec:simu_single_layer}
In this section, we evaluate the performance of the proposed spectral unmixing algorithm on synthetic MSL signals generated using the parameters used in Section \ref{subsec:perf_analysis} (Eq. \eqref{eq:param_exp}) and impulse response approximation \eqref{eq:impulse_resp0}.

The number of spectral bands has been chosen as $L=4,8,16,32$, equally spaced between $400$ nm and $2500$ nm.
For each scenario, the number of iterations has been set to $N_{\textrm{MC}}=8000$ (including $N_{\textrm{bi}}=4000$ burn-in samples).
The hyperparameters have been fixed to $(\alpha^2,\gamma^2)=(10^{6},10^{6})$.
The estimation performance of the proposed algorithm is evaluated by comparing the CRLB defined in Section \ref{sec:CRLB} to the mean square error (MSE) defined as
\begin{eqnarray}
\label{eq:MSE}
    \textrm{MSE}_i= \textrm{E}\left[\left(\hat{\theta}_{i} - \theta_i\right)^2\right], \quad i=1,\ldots,R+L+1
\end{eqnarray} 
where $\theta_i$
and $\hat{\theta}_{i}$ are the actual and estimated $i$th element of the unknown parameter vector $\paramvect$ and the expectation is approximated using $900$ Monte Carlo runs.
The performance of the proposed algorithm is compared to the CRLB (including the Gaussian approximation \eqref{eq:impulse_resp1}) and to the performance of the algorithm developed in \cite{Wallace2014}, denoted as ``State-of-the-art'' and which consists of estimating sequentially the position, amplitudes of the peaks for each wavelength and finally the relative areas.

Table \ref{tab:perf_exp_MSE} shows that the MSEs obtained for $L\in \left\lbrace 4,8,16,32\right \rbrace$ by the two algorithms are close to the CLRBs, although the proposed algorithm generally outperforms the method proposed in \cite{Wallace2014} due to the joint estimation of the target position and the material areas.
Note that for $L=4$, the variance of the proposed estimator is slightly lower than the CRLB.
This can be mainly explained by the fact that the estimator is no longer unbiased in that case.
Table \ref{tab:perf_exp_rel_err} compares the average relative estimation errors obtained by the two algorithms and computed by dividing the square root of the MSEs by the actual values of the parameters. These results show that increasing the number of spectral bands from $L=4$ to $L=32$ almost divides the area estimation errors by three (\emph{e.g.}, from $\approx 30\%$ to $\approx 10\%$ for the bark area), which highlights the benefits of increasing the number of wavelengths.
\begin{table}[h!]
\renewcommand{\arraystretch}{1.2}
\begin{tiny}
\begin{center}
\caption{Estimation performance: MSE
.\label{tab:perf_exp_MSE}}
\begin{tabular}{|c|c|c|c|c|}
\cline{3-5}
\multicolumn{2}{c|}{} &  $w_n$  & $w_b$ & $w_s$  \\
\hline
\multicolumn{2}{|c|}{Actual parameter value} & $0.2$ & $0.3$ & $0.4$ \\
\hline
\hline
\multirow{3}*{$L=32$} & CRLB & $2.6 \times 10^{-4}$& $1.0 \times 10^{-3}$ & $3.6 \times 10^{-5}$ \\
\cline{2-5}
& Proposed Algo. & $ 2.7\times 10^{-4}$& $ 1.1\times 10^{-3}$ & $ 3.8\times 10^{-5}$ \\
\cline{2-5}
 & State-of-the-art & $ 3.0\times 10^{-4}$& $ 1.2\times 10^{-3}$ & $ 4.1\times 10^{-5}$ \\
\hline
\hline
\multirow{4}*{$L=16$} & CRLB & $4.6 \times 10^{-4}$& $1.9 \times 10^{-3}$ & $6.8 \times 10^{-5}$ \\
\cline{2-5}
& Proposed Algo. & $ 4.8 \times 10^{-4}$& $ 2.0\times 10^{-3}$ & $ 7.0 \times 10^{-5}$ \\
\cline{2-5}
 & State-of-the-art & $ 5.3 \times 10^{-4}$ & $ 2.2 \times 10^{-3}$ & $ 7.8 \times 10^{-5}$ \\
\hline
\hline
\multirow{4}*{$L=8$} & CRLB & $5.8 \times 10^{-4}$& $2.6 \times 10^{-3}$ & $1.0 \times 10^{-4}$ \\
\cline{2-5}
& Proposed Algo. & $ 5.7 \times 10^{-4}$& $ 2.5\times 10^{-3}$ & $ 1.0\times 10^{-4}$ \\
\cline{2-5}
 & State-of-the-art & $6.1 \times 10^{-4}$ & $2.7 \times 10^{-3}$ & $1.1 \times 10^{-4}$ \\
\hline
\hline
\multirow{4}*{$L=4$} & CRLB & $1.8 \times 10^{-3}$& $8.0 \times 10^{-3}$ & $3.0 \times 10^{-4}$ \\
\cline{2-5}
& Proposed Algo. & $ 1.6 \times 10^{-3}$& $ 7.4\times 10^{-3}$ & $2.9\times 10^{-4}$ \\
\cline{2-5}
 & State-of-the-art & $1.8 \times 10^{-3}$ & $8.6 \times 10^{-3}$ & $3.3 \times 10^{-4}$\\
\hline
\end{tabular}
\end{center}
\end{tiny}
\vspace{-0.4cm}
\end{table}

\begin{table}[h!]
\renewcommand{\arraystretch}{1.2}
\begin{tiny}
\begin{center}
\caption{Estimation performance: Average relative errors (in \%)
.\label{tab:perf_exp_rel_err}}
\begin{tabular}{|c|c|c|c|c|}
\cline{3-5}
\multicolumn{2}{c|}{} &  $w_n$  & $w_b$ & $w_s$  \\
\hline
\multicolumn{2}{|c|}{Actual parameter value} & $0.2$ & $0.3$ & $0.4$ \\
\hline
\hline
\multirow{3}*{$L=32$} & Theo. (\%) & 8.0 & 10.7 & 1.5 \\
\cline{2-5}
 & Proposed Algo.& 8.2 & 11.0 & 1.5 \\
\cline{2-5}
 & State-of-the-art & 8.6 & 11.5 & 1.6 \\
\hline
\hline
\multirow{3}*{$L=16$} & Theo. (\%) & 10.8 & 14.6 & 2.0 \\
\cline{2-5}
 & Proposed Algo.& 11.0 & 14.8 & 2.1 \\
\cline{2-5}
 & State-of-the-art & 11.6 & 15.7 & 2.2 \\
\hline
\hline
\multirow{3}*{$L=8$} & Theo. (\%) & 12.1 & 16.8 & 2.5 \\
\cline{2-5}
 & Proposed Algo.& 12.0 & 16.8 & 2.5 \\
\cline{2-5}
 & State-of-the-art & 12.4 & 17.4 & 2.6 \\
\hline
\hline
\multirow{3}*{$L=4$} & Theo. (\%) & 21.0 & 29.9 & 4.3 \\
\cline{2-5}
 & Proposed Algo.& 20.0 & 28.8 & 4.2 \\
\cline{2-5}
 & State-of-the-art & 21.4 & 30.8 & 4.5 \\
\hline

\end{tabular}
\end{center}
\end{tiny}
\vspace{-0.4cm}
\end{table}

\subsection{Extension to a multi-layer target}
\label{subsec:multilayer}
\begin{algogo}{Gibbs Sampling Algorithm (multi-layer)}
     \label{algo:bayesian2}
     \begin{algorithmic}[1]
        \STATE \underline{Fixed input parameters:} $\MATmat,\bphi,\alpha^2,\gamma^2$, 
				number of layers $D$, layer positions $\left \lbrace t_d\right \rbrace_{d=1,\ldots, D}$, number of burn-in iterations $N_{\textrm{bi}}$, total number of iterations$N_{\textrm{MC}}$
				\STATE \underline{Initialization ($i=0$)}
        \begin{itemize}
        \item Set $\left \lbrace \bfw_d^{(0)} \right \rbrace_{d=1,\ldots, D},\bfb^{(0)}$
        \end{itemize}
        \STATE \underline{Iterations ($1 \leq i \leq N_{\textrm{MC}}$)}
				\FOR{d=1:D}
        \STATE Sample $\bfw_d^{(i)}$ from its conditional distribution and CHMC
				\ENDFOR
        \STATE Sample $\bfb^{(i)}$ from its conditional distribution and Gaussian random walks
        \STATE Set $i = i+1$.
\end{algorithmic}
\end{algogo}

In this section we extend the model \eqref{eq:model0} by considering a target composed of $D$ layers, leading to 
 \begin{eqnarray}
\label{eq:model1}
\pix{\ell}{\nobin} \sim \mathcal{P}\left(\sum_{d=1}^{D} \Vmat{\ell,:}\bfw_{d}g_{d,\ell}(\nobin) + b_{\ell}\right)
\end{eqnarray}
where $g_{d,\ell}(\cdot)$ is the photon impulse response of the $d$th layer located at $t_d$ and $\bfw_{d}=[w_{1,d},\ldots,w_{R,d}]\transp\succeq 0$ is the relative area vector associated with the $d$th layer. 
In this scenario, we assume that the number and positions of the layers are known. Consequently, the unmixing problem reduces to estimating the relative areas of the $R$ known components for the $D$ layers (and the background levels). The joint spectral and spatial unmixing problem is more challenging and will be discussed in Section \ref{subsec:real_data}. The Bayesian model presented in Section \ref{sec:bayesian} has been extended by considering the same prior \eqref{eq:joint_prior_w} for the area vectors of each layer. The sampling procedure studied in Section \ref{sec:Gibbs} has been modified in order to update sequentially the $D$ area vectors and the target position update step has been removed since the surface positions are assumed to be fixed in this paragraph (see Algo. \ref{algo:bayesian2}).
 
\subsubsection{Target description}
\begin{figure}[h!]
  \centering
  \includegraphics[width=\columnwidth]{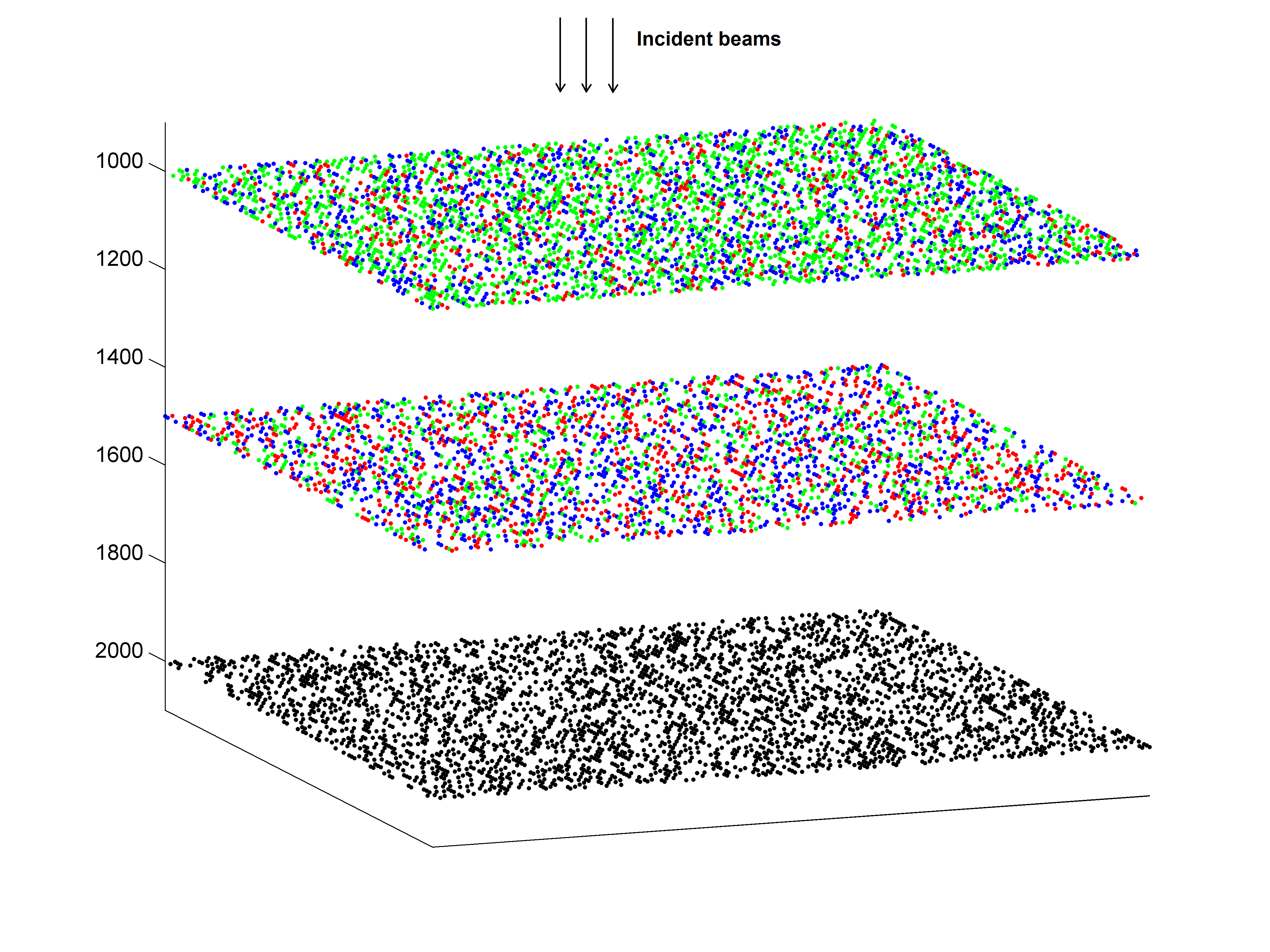}
  \caption{Artificial 3-layered target composed of needles (red), bark (green), soil (blue) and pure reflective material (black).}
  \label{fig:3layers_scene}
\end{figure}

\begin{figure}[h!]
  \centering
  \includegraphics[width=\columnwidth]{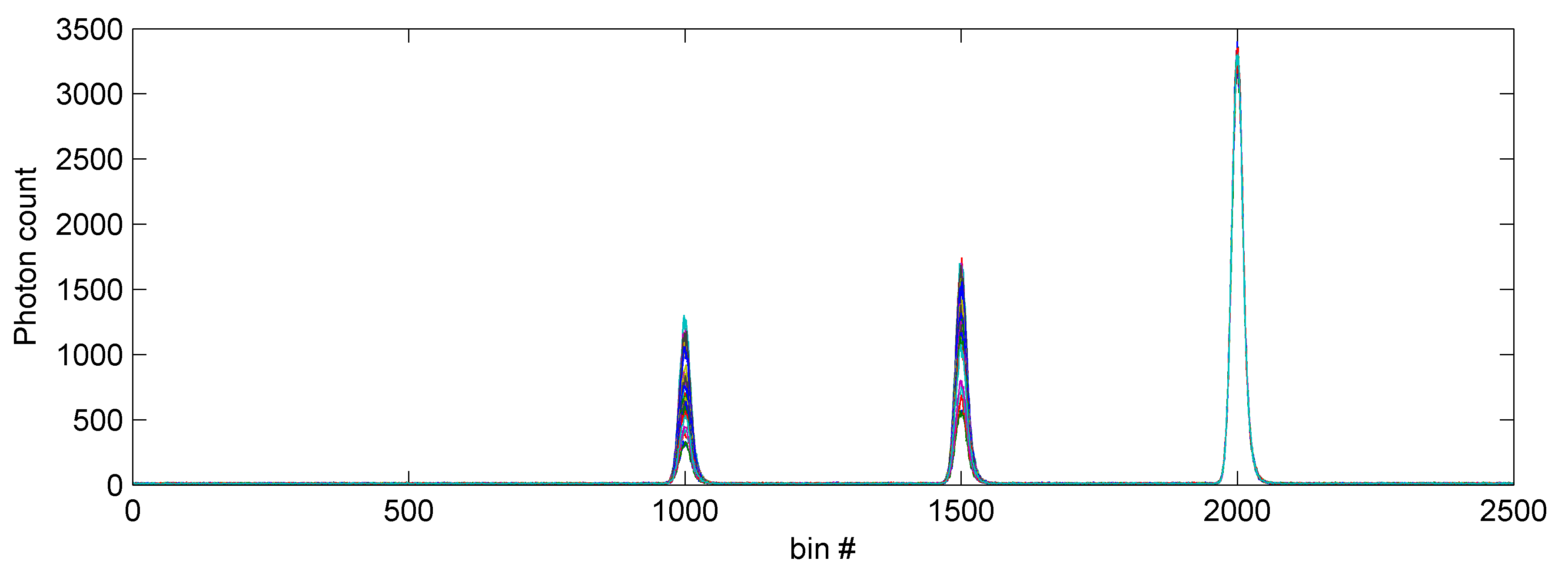}
  \caption{MSL signals ($L=32$) associated with the artificial 3-layered target ($t_1=1000,t_2=1500,t_1=2000$). The different colors correspond to different spectral bands.}
  \label{fig:MSL_3layers}
\end{figure}

We evaluate the unmixing performance of the extended Bayesian algorithm using an artificial target composed of $D=3$ layers. More precisely, the multi-layer target is assumed to be far enough from the source to ensure that the laser rays hitting the target have the same incident direction. Thus the photons reflected onto the $d$th layers lead to the same impulse response (\emph{i.e.}, same $t_d$). The first two layers are located at $t_1=1000$ and $t_2=1500$ and are composed of needles, bar and soil. It is important to note that the $3$ materials composing the first two layers are assumed to randomly distributed, without overlap.  The third layer modeling the reference spectralon is located at $t_3=2000$. The associated relative areas are presented in Table \ref{tab:perf_3layers} and the location of the $D=3$ layers is depicted in Fig. \ref{fig:3layers_scene}. Note that in this particular scenario, the relative areas correspond to areas visible by the source (no occlusion) and satisfy $\sum_{r=1}^{R}\sum_{d=1}^{D}w_{r,d}=1$. However, this constraint is not ensured by the proposed estimation procedure. MSL data associated with this scene have been generated according to \eqref{eq:model1} with $\beta=10^{4}, b_{\ell}=b=10, \forall \ell$ and $L=32$ and are depicted in Fig. \ref{fig:MSL_3layers}
% (\red{the relative areas used to generate the data are not the actual one but areas associated with virtual ray tracer which samples uniformly the area of interest. To be discussed}).  

\subsubsection{Estimation performance}
The proposed Bayesian algorithm extended to account for multiple layers has been applied to the MSL signals with $N_{\textrm{MC}}=8000$ (including $N_{\textrm{bi}}=4000$ burn-in samples). The layer positions have been fixed to their actual values. Table \ref{tab:perf_3layers} shows the estimated relative areas for each layer and $L\in \left\lbrace 4,8,16,32\right \rbrace$.
This table shows that the proposed algorithm provides accurate area estimates for each layer and the more wavelengths the better the estimation, as already observed with the single-layered target in Section \ref{subsec:simu_single_layer}.

\begin{table}[h!]
\renewcommand{\arraystretch}{1.2}
\begin{small}
\begin{center}
\caption{Estimation performance: Multi-layer target.\label{tab:perf_3layers}}
\begin{tabular}{|c|c|c|c|c|c|}
\cline{3-6}
\multicolumn{2}{c|}{} &  $w_n$  & $w_b$ & $w_s$ & spectralon \\
\hline
\multirow{3}*{Actual} & Layer 1 & $0.099$& $0.099$ & $0.102$ & $0$ \\
\cline{2-6}
&Layer 2 & $ 0.080$& $ 0.200$ & $0.120$ & $0$ \\
\cline{2-6}
 & Layer 3 & $0$ & $0$ & $0$ & $0.30$  \\
\hline
\hline
\multirow{3}*{$\underset{(L=32)}{\textrm{Estimated}}$} & Layer 1 & $0.0928$& $0.0780$ & $0.1185$ & $0.0004$\\
\cline{2-6}
&Layer 2 & $ 0.0669$& $ 0.1953$ & $ 0.1161$ & $0.0006$\\
\cline{2-6}
 & Layer 3& 0.0019 & 0.0011 & 0.0011 & $0.3274$\\
\hline
\hline
\multirow{3}*{$\underset{(L=16)}{\textrm{Estimated}}$} & Layer 1 & $0.0932$& $0.0773$ & $0.1181$ & $0.0006$\\
\cline{2-6}
&Layer 2 & $ 0.0620$& $ 0.1961$ & $0.1179$ & $0.0006$\\
\cline{2-6}
 & Layer 3& $0.0024$ & $0.0018$ & $0.0014$ & $0.3270$\\
\hline
\hline
\multirow{3}*{$\underset{(L=8)}{\textrm{Estimated}}$} & Layer 1 & $0.0957$& $0.0764$ & $0.1166$& $0.0008$ \\
\cline{2-6}
&Layer 2 & $ 0.0640$& $ 0.1954$ & $ 0.1172$& $0.0006$ \\
\cline{2-6}
 & Layer 3& $0.0024$ & $0.0028$ & $0.0016$ & $0.3268$\\
\hline
\hline
\multirow{3}*{$\underset{(L=4)}{\textrm{Estimated}}$} & Layer 1 & $0.1286$& $0.0687$ & $0.1013$ & $0.0015$\\
\cline{2-6}
&Layer 2 & $0.0975$& $ 0.1863$ & $ 0.1014$ & $0.0016$\\
\cline{2-6}
 & Layer 3 & $0.0023$ & $0.0019$ & $0.0015$ & $0.3271$\\
\hline
\end{tabular}
\end{center}
\end{small}
\vspace{-0.4cm}
\end{table}

\subsection{Towards real data analysis}
\label{subsec:real_data}

Due to the current limitations of the instrument, the proposed algorithm has only been applied to synthetic MSL signals whose underlying model has been shown to be in good agreement with our previous real MSL measurements using fewer wavelengths~\cite{Hernandez2007,Wallace2014}.
Using this real data, we have managed to investigate the precision of area estimation in a multi-layer model; ground truth was available for the endmembers, but unfortunately we did not have structural truth.
As stated above, real MSL signal acquisition campaigns are underway with $3$ wavelengths~\cite{Fleming2015} but we need to extend these and have simultaneous ground truth measurements to better evaluate the accuracy of the model and the estimation algorithm presented in this paper.
Moreover, the number and positions of the layers which the multi-layer target is composed of were assumed to be known in Section \ref{subsec:multilayer}.
Estimating these parameters, especially the number of layers, for MSL data is a 
challenging problem that can be addressed in the Bayesian framework using reversible jump MCMC methods, in a similar fashion to \cite{Hernandez2007}. 
This issue is however out of scope of this work and will also be addressed in a future paper.

\section{Conclusion}
\label{sec:conclusion} 

In this paper, we have proposed and developed a Bayesian model and an MCMC method for spectral unmixing of multispectral Lidar data.
First, we evaluated our method on a single layer simulated target composed of known materials
so that the MSL returns consisted of the sum of the individual contributions of the different components. The algorithm estimated the target position, the relative areas of the materials and the noise statistical properties.

The Cramer-Rao lower bound associated with the observation model was derived to identify the key parameters involved in the expected performance of the SU algorithm.
It was shown that this bound can be used to help design future multispectral/hyperspectral Lidar systems (\emph{e.g.,} number of spectral bands and associated wavelengths, laser power \ldots) for specific applications. It was also shown that the performance of the proposed SU strategy is close to the results provided by the Cramer-Rao lower bound, although the bound considered in the paper only holds for deterministic parameters. The proposed Bayesian model was then extended to handle multi-layer targets (assuming the layer positions are known) and the simulation results provided interesting results for solving the more challenging joint spectral and spatial unmixing problem.

The model use in this paper does not take into account possible multiple scattering effects (which are likely to occur when analyzing multi-layer scenes).
Such effects have already been observed in HSIs and Lidar signals over canopies.
Developing non-linear models for MSL signal analysis is a challenging problem to be addressed in future studies.

\section*{Appendix: Fisher information matrix}
\label{sec:appendix}
The likelihood of the observation matrix $\MATpix$ can be expressed as 
\begin{eqnarray}
f(\MATpix| \bfw,\MATmat,t_0,\bfb) = \prod_{t=1}^{T}\prod_{\ell}^{L} 
\dfrac{\left(\lambda_{\ell,t} \right)^{\pix{\ell}{t}}}{\pix{\ell}{t}!}
\exp^{-\lambda_{\ell,t}}
\end{eqnarray}
where $\lambda_{\ell,t}=\Vmat{\ell,:}\bfw \tilde{g}_{0,\ell}(\nobin) + b_{\ell}$. The corresponding log-likelihood $P= \log f(\MATpix| \bfw,\MATmat,t_0,\bfb)$ is given by 
\begin{eqnarray}
P  = \sum_{t=1}^{T}\sum_{\ell=1}^{L} 
\pix{\ell}{t} \log \left(\lambda_{\ell,t}\right) - \log \left(\pix{\ell}{t}!\right)  -\lambda_{\ell,t}.
\end{eqnarray}
The partial derivatives of $P$ with respect to (w.r.t.) the unknown model parameters are 
\begin{eqnarray}
\dfrac{\partial P}{\partial \bfw}  & = &  \sum_{\ell=1}^{\nbband} \sum_{\nobin=1}^{\nbbin} \left(\dfrac{\pix{\ell}{\nobin}\tilde{g}_{0,\ell}(t)}{\lambda_{\ell,t} } -\tilde{g}_{0,\ell}(\nobin)\right) \Vmat{\ell,:}\nonumber\\
\dfrac{\partial P}{\partial b_{\ell}}  & = & \sum_{\nobin=1}^{\nbbin} {\dfrac{\pix{\ell}{\nobin}}{\lambda_{\ell,t}} - 1}\nonumber\\
\dfrac{\partial P}{\partial t_0}  & = & \sum_{\nobin=1}^{\nbbin} \sum_{\ell=1}^{L}{\left(\dfrac{\pix{\ell}{\nobin}}{\lambda_{\ell,t}} - 1\right)\Vmat{\ell,:}\bfw \tilde{g}_{0,\ell}(\nobin)\dfrac{t-t_0}{\sigma^2}}\nonumber.
\end{eqnarray}
Straightforward computations lead to 
\begin{eqnarray}
\textrm{E} \left[\dfrac{\partial^2 P}{\partial \bfw^2} \right]& = & - \sum_{\nobin=1}^{\nbbin} \sum_{\ell=1}^{\nbband} \dfrac{\tilde{g}_{0,\ell}^2(\nobin)}{\lambda_{\ell,t}}\Vmat{\ell,:}\transp\Vmat{\ell,:}\nonumber\\
\textrm{E} \left[\dfrac{\partial^2 P}{\partial b_{\ell}\partial b_{\ell'}}\right] & = & \left\{
    \begin{array}{ll}
        - \sum_{z=1}^{N_z} \dfrac{1}{\lambda_{\ell,t}}& \mbox{if } \ell=\ell' \\
				0 & \mbox{else }\\
    \end{array}
\right.\nonumber\\
\textrm{E} \left[\dfrac{\partial^2 P}{\partial \bfw\partial b_{\ell}}\right] & = & -\sum_{\nobin=1}^{\nbbin} \dfrac{\tilde{g}_{0,\ell}(\nobin)}{\lambda_{\ell,t} }\Vmat{\ell,:}\nonumber\\
\textrm{E} \left[\dfrac{\partial^2 P}{\partial t_0^2}\right] & = & -\sum_{\nobin=1}^{\nbbin}\sum_{\ell=1}^{\nbband} \dfrac{\left[(\nobin-t_0)\Vmat{\ell,:}\bfw \tilde{g}_{0,\ell}(\nobin)\right]^2}{\sigma^4\lambda_{\ell,t}}\nonumber\\ 
\textrm{E} \left[\dfrac{\partial^2 P}{\partial t_0\partial \bfw}\right] & = & -\sum_{\nobin=1}^{\nbbin}\sum_{\ell=1}^{\nbband} \dfrac{(\nobin-t_0)\Vmat{\ell,:}\bfw \tilde{g}_{0,\ell}^2(\nobin)}{\sigma^2\lambda_{\ell,t}}\Vmat{\ell,:}\nonumber\\
\textrm{E} \left[\dfrac{\partial^2 P}{\partial t_0\partial b_{\ell}}\right] & = & -\sum_{\nobin=1}^{\nbbin} \dfrac{(\nobin-t_0)\Vmat{\ell,:}\bfw \tilde{g}_{0,\ell}(\nobin)}{\sigma^2\lambda_{\ell,t}}.\nonumber 
\end{eqnarray}
using the fact fact $\textrm{E} \left[\pix{\ell}{\nobin} \right]=\Vmat{\ell,:}\bfw \tilde{g}_{0,\ell}(\nobin) + b_{\ell}$
\bibliographystyle{IEEEtran}
\bibliography{biblio}
\end{document}